\documentclass[pss,fleqn]{w-art}
\usepackage{times}
\usepackage{w-thm}
\usepackage[]{graphicx}
\begin{document}
\DOIsuffix{theDOIsuffix}
\Volume{XX}
\Issue{1}
\Month{01}
\Year{2003}
\pagespan{1}{}
\Receiveddate{}
\Reviseddate{}
\Accepteddate{}
\Dateposted{}



\title[Quasiclassical theory of Luttinger liquids]{Quasiclassical theory of
electronic transport in mesoscopic systems: Luttinger liquids
revisited}


\author[U. Eckern]{Ulrich Eckern\footnote{Corresponding
     author: e-mail: {\sf ulrich.eckern@physik.uni-augsburg.de}}} 
     \address[]{Institut f\"ur Physik, Universit\"at Augsburg, 86135
                Augsburg, Germany}
\author[P. Schwab]{Peter Schwab\footnote{E-Mail: 
        {\sf peter.schwab@physik.uni-augsburg.de}}}
\begin{abstract}
The method of the quasiclassical Green's function is used to determine
the equilibrium properties of one-dimensional (1D)
interacting Fermi systems, in particular, the bulk and the local (near
a hard wall) density of states.
While this is a novel approach to 1D systems, our findings do 
agree with standard results for Luttinger liquids obtained
with the bosonization method.
Analogies to the so-called $P(E)$ theory of tunneling through
ultrasmall junctions are pointed out and are exploited.
Further applications of the Green's function method for 1D systems are
discussed.
\end{abstract}
\maketitle 




\section{Introduction}
Quasiclassical methods, developed several years ago and used
successfully
to describe non-equilibrium states in superconductors and superfluids
(see,
e.g., \cite{schmid1975,larkin1976,eckern1981,eckern1981a}), 
have recently been extended to meso- and nanoscopic systems.
This became possible after the formulation of boundary conditions,
including hybrid structures and interface roughness
\cite{shelankov2000,luck2001,luck2004}.
In recent developments  we focussed on spin-effects and
spatial confinement, e.g., concerning the spin-Hall effect in a 2D
electron gas \cite{raimondi2006}, and spin relaxation in narrow wires 
in the presence of spin-orbit coupling \cite{schwab2006}.
Within the quasiclassical approach, it is also possible to study the influence of
disorder and Coulomb interaction on the same footing 
\cite{kamenev1999,schwab2003}, which is manageable since the
method works on an intermediate level: microscopic details of the
wave-functions, on the scale of the interatomic distance, are
integrated out,
leaving equations of motion which can be solved with sufficient
accuracy.  

As an illustrative example, we apply in this paper 
the quasiclassical method to
determine the equilibrium Green's function, and hence the density of
states (DoS), of the (spinless) 1D model known as 
Luttinger liquid \cite{tomonaga1950,luttinger1963, haldane1981}. This
model contains two fermionic branches, linearized near the Fermi points,
with $v_F$ and $N_0 = (\pi v_F)^{-1}$ the bare Fermi velocity and DoS.
Only interaction processes with small ($\ll k_F$) momentum transfer are
considered: standard parameters are $g_4$ and $g_2$, for scattering
processes involving only one or both branches, respectively. The
dimensionless quantities $\gamma_4 = g_4/2\pi v_F$ and
$\gamma_2 = g_2/2\pi v_F$ are useful. The spectrum of charge fluctuations
is linear, $\omega (q) = v |q|$; the renormalized velocity, $v$, and
the parameter $K$, given by
\begin{equation}
\label{vK}
v = v_F \left[ (1 + \gamma_4)^2 - \gamma_2^2 \right]^{1/2}
\; , \;\;
K =  \left[ \frac{1+\gamma_4-\gamma_2}{1+\gamma_4+\gamma_2} \right]^{1/2}
\; ,
\end{equation}
are often called Luttinger liquid parameters.

In the following Sect.\ 2 we discuss the solution of the equation of
motion for the quasiclassical Green's function in the presence of a
fluctuating potential, the latter representing the fermion-fermion
interaction. The Keldysh technique is employed
throughout. Considering the average -- with respect to the fluctuating
field -- of the Green's function and an analogy to the $P(E)$ theory,
the field fluctuations are related to an effective impedance of the
Luttinger liquid, and finally to the DoS (Sect.\ 3). In Sect.\ 4 we
consider an insulating boundary, i.e.\ a reflecting wall, and determine
the DoS boundary exponent. A concluding discussion is given in 
Sect.\ 5.

\section{Quasiclassical equation of motion and its solution}
In essence, the Keldysh technique \cite{keldysh1965,rammer1986}
differs from the zero-temperature and
the Matsubara approach by employing time-ordering along a contour which
runs from $-\infty$ to $+\infty$ and back to $-\infty$. Rewriting the 
theory in terms of standard ($-\infty ... +\infty$) integrations, a
2$\times$2 matrix structure results; e.g., the Green's function can be
cast into the form
\begin{equation}
\label{G}
\check{G} = \left\{ \begin{array}{cc}
            G^R & G^K \\ 0 & G^A \end{array} \right\}
\end{equation}
Accordingly, considering some general 2-particle interaction $V_0$ and
decoupling this interaction using the Hubbard-Stratonovich transformation
within the path-integral formalism, two auxiliary fields are necessary.
As a result, the Green's function can be expressed as an average of the
free-particle Green's function in the presence of fluctuating potentials
as follows (see, e.g., Ref.\ \cite{kamenev1999}):
\begin{equation}
\label{av}
\check{G} = \langle \check{G}_0 [\Phi] \, {\rm e}^{{\rm Tr}\ln (\check{1} +
\check{G}_0 \check{\Phi} )} \rangle_{0, \Phi }
= \langle \check{G}_0 [\Phi] \rangle_\Phi
\end{equation}
Here $\check{G}_0$ is the free-particle Green's function, and
$\check{G}_0 ^{-1}[\Phi] = \check{G}_0^{-1} + \check{\Phi}$, where
$\check{\Phi} = \phi_1 \check{\sigma}_0 + \phi_2 \check{\sigma}_1$;
$\check{\sigma}_0$ and $\check{\sigma}_1$ denote the unit and the first
Pauli matrix, and $\phi_1$ and $\phi_2$ correspond to the two potentials
mentioned above. Furthermore, $\langle ...\rangle_{0,\Phi}$ denotes an
average which is Gaussian by construction,
and defined such that
\begin{equation}
\label{phi}
\langle \phi_i (xt) \phi_j (x^\prime t^\prime) \rangle_{0, \Phi} =
\frac{i}{2}
\left\{ \begin{array}{cc}
V_0^K & V_0^R \\ V_0^A & 0  \end{array} \right\}_{ij} (xt,x^\prime t^\prime)
\end{equation}
where for clarity space ($x$) and time ($t$) arguments are included. 
The average $\langle... \rangle_\Phi$, defined through
Eqs.~(\ref{av}) and (\ref{phi}), can be non-Gaussian.
(The spin will be suppressed througout this paper.) Note that a ``real''
electrical potential can be included by the replacement 
$\phi_1 \to \phi_1 + e\phi_{\rm ext}$ in $\check{G}_0 [\Phi]$ of Eq.\
(\ref{av}); the particle's charge is $-e$. In the next step, we consider
the quasiclassical approximation, i.e.\ we consider the difference
$\check{G}_0^{-1} [\Phi] \check{G}_0 [\Phi] - 
\check{G}_0 [\Phi] \check{G}_0^{-1} [\Phi] = 0$, keeping only the
leading terms in the gradients with respect to the (spatial)
center-of-mass coordinate (which we again denote by $x$). 
The equation is then integrated with respect to the magnitude of the
momentum.
The result is
\begin{equation}
\label{equ}
\left[ \frac{\partial}{\partial t} + \frac{\partial}{\partial t^\prime}
       + v_F \hat{p}\cdot\nabla_x \right] \check{g}_{tt^\prime}(\hat{p}x)
       = i [\check{\Phi}, \check{g}]
\end{equation}
where $\hat{p}$ is the direction of the center-of-mass momentum, and
\begin{equation}
\label{comm}
[\check{\Phi}, \check{g}] \equiv 
\check{\Phi}(xt)\check{g}_{tt^\prime}(\hat{p}x) - 
\check{g}_{tt^\prime}(\hat{p}x) \check{\Phi}(xt^\prime) \; .
\end{equation}
The quantity $\check{g}_{tt^\prime}(\hat{p}x)$ is the so-called
quasiclassical Green's function, 
\begin{equation}
\check g_{tt'}(\hat p x) = \frac{i}{\pi} \int d \xi \check G_{tt'}({\bf p},
x), \quad \xi = p^2/2m - \mu
;\end{equation}
a major advantage compared to the full Green's function is the normalization,
$\check{g}\check{g} = \check{1}$. It should be noted that in
the quasiclassical approximation, only long-wavelength contibutions of
the fluctuating fields are taken into account -- which is adequate for
the problem addressed below. In the 1D case, $\nabla_x \to \partial_x$,
and $\check{p} = \pm$. For the homogeneous equilibrium case, the 
solution is
\begin{equation}
\label{gnull}
\check{g}_0 (\epsilon) =
\left\{ \begin{array}{cc}
1 & 2 F(\epsilon) \\ 0 & -1 \end{array} \right\}
\end{equation}
where the Fourier transformed ($t - t^\prime \to \epsilon$) quantity
is given, and $F(\epsilon) = \tanh (\beta\epsilon /2)$, $\beta = 1/k_B T$. 
A formal solution of Eq.\ (\ref{equ}) is found with the ansatz
\begin{equation}
\label{ansatz}
\check{g}_{tt^\prime}(\hat{p}x) = {\rm e}^{i\check{\varphi}(xt)}
\, \check{g}_{0,t-t^\prime} \, {\rm e}^{-i\check{\varphi}(xt^\prime)}
\end{equation}
provided
\begin{equation}
\label{solution}
(\partial_t + v_F \hat{p} \partial_x ) \check{\varphi}(xt)
\equiv D_{xt} \check{\varphi}(xt)
= \check{\Phi} (xt) \; .
\end{equation}
Apparently, $\check{\varphi}$ has the same matrix structure as
$\check{\Phi}$, namely $\check{\varphi} = \varphi_1 \check{\sigma}_0 +
\varphi_2 \check{\sigma}_1$. For example, we obtain explicitly from
(\ref{ansatz}):
\begin{equation}
g^R (tt^\prime) = \delta (t-t^\prime) -2i F (tt^\prime)
  \cos\varphi_2 \sin\varphi_2^\prime \, 
  {\rm e}^{i(\varphi_1 -\varphi_1^\prime)}
\end{equation}
\begin{equation}
g^K (tt^\prime) = 2 F (tt^\prime) \cos\varphi_2 \cos\varphi_2^\prime \, 
{\rm e}^{i(\varphi_1 -\varphi_1^\prime)}
\end{equation}
where $\varphi_j \equiv \varphi_j (xt)$ and 
$\varphi_j^\prime \equiv \varphi_j (xt^\prime)$ for brevity. We will discuss
below that, for the Luttinger model, it is sufficient to assume that the
phases are Gaussian distributed. It is then straightforward to determine
$\langle \check{g} \rangle_\phi$; for example, we obtain\footnote{The
quantities discussed from here on depend on the time difference, 
$t-t^\prime$, but we nevertheless use the notation $tt^\prime$ 
for brevity.}
\begin{eqnarray}
\langle \check{g}^R \rangle_\Phi (tt^\prime) & = & \delta (t-t^\prime) 
+  F (tt^\prime) \, {\rm e}^{- J_0 (tt^\prime)} \,
[ \sinh (J_1(tt^\prime)) - \sinh (J_2(tt^\prime)) ] \\
   & = & - \langle \check{g}^A \rangle_\Phi (t^\prime t)
\end{eqnarray}
where
\begin{eqnarray}
J_0 (tt^\prime) = & \langle (\varphi_1 - \varphi_1^\prime)^2 \rangle_\Phi /2 
                 & =  J^K (0) - J^K (tt^\prime) \; , \\
J_1 (tt^\prime) = & \langle (\varphi_1 - \varphi_1^\prime)
    (\varphi_2 + \varphi_2^\prime) \rangle_\Phi 
                 & =  J^R (tt^\prime) - J^A (tt^\prime) \; ,\\
J_2 (tt^\prime) = & \langle (\varphi_1 - \varphi_1^\prime) 
    (\varphi_2 - \varphi_2^\prime) \rangle_\Phi
                 & =  - J^R (tt^\prime) - J^A (tt^\prime) \; .
\end{eqnarray}
Note that $J_0$ and $J_2$ are even, and $F$ and $J_1$ are odd under time
reversal, $t-t^\prime \to t^\prime -t$. Then we define $J=-J_0+J_1$,
such that
\begin{equation}
\label{J}
J(t) = \int \frac{d\omega}{2\pi} \left[ J^R(\omega) - J^A (\omega) \right]
       \left[ B(\omega) + 1 \right] \left( {\rm e}^{-i\omega t} -1 \right)
\end{equation}
where $B(\omega) = \coth (\beta\omega/2)$, and we used
$J^K (\omega) = [ J^R(\omega) - J^A (\omega) ] B (\omega)$. From 
the relation $J^R (tt^\prime) = J^A (t^\prime t)$ we obtain
$J^A (\omega) = J^R (-\omega)$, implying that $J^R(\omega)-J^A(\omega)$
is odd in frequency.
Combining the above relations, the (normalized) DoS is given by
\begin{eqnarray}
\label{DoS}
N(\epsilon)\, =  \, {\rm Re} g^R(\epsilon) & = & 1 + \pi \int dt \, {\rm e}^{i\epsilon t} \,
  \{ F(tt^\prime) [P(tt^\prime) - P(t^\prime t)]\}_{t^\prime=0} 
  \nonumber \\
  & = & 1 + \frac{1}{2} \int d\omega \, F(\epsilon - \omega)
[P(\omega) - P(-\omega)]
\end{eqnarray}
with
\begin{equation}
\label{P}
P(\omega) = \frac{1}{2\pi} \int dt \, {\rm e}^{J(t) +i\omega t} 
.\end{equation}
Note that $P(t=0) = 1/2\pi$, i.e.\ $\int d\omega P(\omega) = 1$.

In order to complete the argument, the phase fluctuations are
easily related to the potential fluctuations, according to the relation
(compare Eq.\ (\ref{solution}))
\begin{equation}
J^R (t) = \langle \varphi_1 (xt) \varphi_2 (x0) \rangle_\Phi
        =  [ D_{xt}^{-1} D_{x^\prime t^\prime}^{-1} \langle \phi_1 (xt) 
              \phi_2 (x^\prime 0) \rangle_\Phi ]_{x=x^\prime,t^\prime =0} \; ,
\end{equation}
and the potential fluctuation can be related to the interaction. This 
last step, however, requires some discussion. First, note that expanding
the exponent in Eq.\ (\ref{av}) up to second order, neglecting higher
order terms, corresponds to the random phase approximation (RPA). Given
this (Gaussian) approximation, the procedure for performing the phase
average, outlined above, is permitted. As an important point, however,
{\em the RPA is known to be exact for calculating the density response
of the Luttinger model}, and hence the effective interaction, which
concludes the argument: the potential fluctuations are given by
\begin{equation}
\langle \phi_1 (xt) \phi_2 (x^\prime t^\prime) \rangle_\Phi =
\frac{i}{2} V^R (xt,x^\prime t^\prime)
\end{equation}
where $V^R (xt,x^\prime t^\prime)$ is the screened (RPA) interaction.
Thus $J^R = (i/2) D^{-1} D^{-1} V^R$ in an obvious short-hand notation.

\section{Effective impedance and bulk DoS}
The approach described in the preceding section is very similar to
the theory of charge tunneling in ultrasmall junctions, as reviewed,
e.g., in \cite{ingold1992}, which is also known as $P(E)$ theory.
(See also \cite{sassetti1994} for related articles.)
In this context, the quantity $P(E)$ characterizes the influence of
the environment on the tunneling between the two electrodes. For
example, the forward tunneling rate is found to be given by
\begin{equation}
\label{forward}
{\vec \Gamma} (V) = (e^2 R_T)^{-1} \int dE dE^\prime \,
           f(E) [ 1 - f(E^\prime + eV) ] \, P(E - E^\prime )
\end{equation}
where $V$ is the voltage, $R_T$ the tunneling resistance, and $f(E)$ 
the Fermi function. Note the detailed balance condition 
$P(-E) = {\rm e}^{-\beta E} \, P(E)$. Physically, $P(E)$ describes the
probability of exchanging the energy $E$ with the environment
\cite{ingold1992}.

On the other hand, in the present context, consider 
the tunneling rate from the interacting one-dimensional wire into 
a non-interacting lead, which clearly is given by
\begin{equation}
\label{forwardb}
{\vec \Gamma} (V) = (e^2 R_T)^{-1} \int d\epsilon 
           f(\epsilon)N(\epsilon)[ 1 - f(\epsilon + eV) ] 
\end{equation}
with $N(\epsilon)$ the normalized DoS as given in Eq.\ (\ref{DoS}).
Using this equation as well as the properties of $P(\omega)$ as 
described in the previous section and the detailed balance condition,
it is straightforward to confirm that Eqs.\ ({\ref{forward}}) and 
(\ref{forwardb}) coincide, provided $P(E)$ from the tunneling theory
is identified with $P(\omega)$. The reason for the complete
correspondence between these two quantities is apparent on physical
grounds, since the quantity $P(\omega)$ characterizing the Luttinger
liquid arises from the auxiliary potential fluctuations due to the 
fermion-fermion interaction. In the zero-temperature limit, one
finds easily
\begin{equation}
\label{T=0}
N(\epsilon) = \int_0^{|\epsilon|} d\omega P(\omega)
     \quad\quad (T=0) \; .
\end{equation}
Exploiting the analogy further, we identify
\begin{equation}
\label{Z}
J^R(\omega)-J^A(\omega) \equiv 
   \frac{2\pi}{\omega} \frac{{\rm Re}Z(\omega)}{R_K}
\end{equation}
where $R_K = h/e^2 = 2\pi\hbar /e^2$ is the Klitzing constant, and
we may call $Z(\omega)$ effective impedance of the Luttinger liquid.
For the simple example of an ohmic impedance with a high-frequency
cut-off
\begin{equation}
\label{ohmic}
\frac{{\rm Re}Z(\omega)}{R_K} = \frac{1}{g}
     \frac{1}{1 + (\omega/\omega_R)^2}
\end{equation}
the result is \cite{ingold1992}
\begin{equation}
P(\omega) = \frac{{\rm e}^{-2\gamma /g}}{\Gamma (2/g)}
            \frac{1}{\omega} \left[ 
            \frac{\omega}{\omega_R} \right]^{2/g} 
            \quad\quad (T=0, \,     0<\omega < \omega_R)
\end{equation}
where $\gamma$ is Euler's constant.
Thus the DoS is found to vanish at the Fermi energy,
$N(\epsilon) \sim |\epsilon|^{2/g}$, where the exponent is given by
\begin{equation}
\label{2overg}
\frac{2}{g}  =  2 \cdot \left\{ \frac{\omega}{2\pi} 
                [J^R (\omega)-J^A (\omega)]\right\}_{\omega\to 0} 
             =  -i \left\{ \frac{\omega}{\pi} \int \frac{dq}{2\pi}
     \frac{V^R(q\omega)}{(-i\omega +0 + i q v_F)^2} \right\}_{\omega\to 0}
\end{equation}
where $V^R(q\omega)$ is the screened retarded interaction. In order to
determine this quantity, consider the 2$\times$2 matrix structure of
right- and left-moving particles, denoted by ``$+$'' and ``$-$'', the
bare interaction and the non-interacting response function:
\begin{equation}
\label{V0}
\hat{V}_0 = \left\{ \begin{array}{cc}
            g_4 & g_2 \\ g_2 & g_4 \end{array} \right\} \; , \;\;
\hat{\chi}_0 = \frac{1}{2\pi v_F} \left\{ \begin{array}{cc}
           \frac{q v_F}{-\omega - i0 + q v_F} & 0 \\
           0 & \frac{q v_F}{+\omega + i0 + q v_F} \end{array} \right\} \; ,
\end{equation}
as well as the RPA equation
$\hat{V}^R = (\hat{1} + \hat{V}_0 \hat{\chi}_0 )^{-1} \hat{V}_0$. 
This matrix structure implies that we have to introduce a branch index
for the fluctuating fields, which we suppressed up to now:
$\check \varphi, \check \Phi \to \check \varphi_\pm, \check \Phi_\pm $.
The density of states of the right-moving particles, for example, is 
determined from the $\check\varphi_+$-$\check\varphi_+$ correlation
functions.
The relevant quantity, $\hat{V}^R_{++}$, is straightforwardly 
determined and inserted into Eq.\ (\ref{2overg}); the $q$-integral 
can be done by contour integration, and we obtain the standard 
\cite{larkin1974} result:\footnote{See \cite{oreg1996} for a recent
summary. Note that often the parameter $K$ is denoted by $g$, which we
avoid here since we prefer to use the latter symbol for the dimensionless
conductance, see Eq.\ (\ref{ohmic}), in accordance with the $P(E)$
theory \cite{ingold1992}.}
\begin{equation}
\label{exponent1}
2/g = (K + K^{-1} -2)/2 \; .
\end{equation}

\section{DoS boundary exponent}
Near a boundary or an interface the quasiclassical approximation is
insufficient, and
the quasiclassical propagators pointing into or out of the boundary (or
interface) have to be connected by boundary conditions
\cite{zaitsev1984,shelankov2000}.
The boundary condition is particularly simple for spinless fermions
in one dimension at an impenetrable wall: obviously 
$\check g_{tt'}(\hat p x) = \check g_{tt'}(-\hat p x )$, to ensure
that there is no current through the wall.

In the following we assume that the wall is located at $x=0$, and that the
particles move in the half-space $x<0$.
Instead of considering two branches of fermions in this
half-space,
we find it more convenient to mirror the left movers at the boundary and to
consider only right movers in the full space, i.e.\
\begin{eqnarray}
\check g_{tt'}(+,x)& = &\left\{  
\begin{array}{ll}
\check g_{tt'}(+,x) & \text{ for } x< 0  \\
\check g_{tt'}(-,-x) & \text{ for } x> 0
\end{array}
\right. \\
\check \Phi_+(x)   & = &
\left\{
\begin{array}{ll}
\Phi_+( x ) & \text{ for } x< 0  \\
\Phi_-( x ) & \text{ for } x> 0
\end{array}
\right.
\end{eqnarray}
This Green's function solves Eq.~(\ref{equ}) both for $x \geq 0$
and $x \leq 0$. The bare interaction is now given in real space by
\begin{equation}
V_0(x,x') = g_4 \delta(x-x') + g_2  \delta(x+x') \;
.\end{equation}
The Fourier transfrom of $V_0$ is thus off-diagonal in momentum space,
since the $g_2$-term couples $q$ with $-q$.
Considering now the 2$\times$2 matrix structure of particles with momentum
$q$ and $-q$, the bare interaction is
\begin{equation}
 \left\{ \begin{array}{cc}
 V_{0,q,q}            & V_{0,q,-q} \\ 
 V_{0,-q,q} & V_{0,-q,-q} \end{array} \right\} 
 = \left\{ \begin{array}{cc}
            g_4 & g_2 \\ g_2 & g_4 \end{array} \right\}  \;
,\end{equation}
i.e.\ identical 
to what was given in Eq.~(\ref{V0}) above with just a different meaning
of the matrix index.
Also the non-interacting response function and  
the RPA equation for the screended interaction are the same as
before.
Due to the absence of translation symmetry 
the effective impedance depends on the distance from the boundary, and
is given by
\begin{eqnarray}
\frac{{\rm Re} Z(x, \omega) }{R_K }
& =  &\frac{\omega}{2 \pi } {\rm Im} 
\left[
\int \frac{dq}{2\pi} \frac{V^R_{q,q}(\omega)}{(-i\omega + i q v_F)^2} 
                    +e^{2 i q x} 
                    \frac{V^R_{q,-q}(\omega)} 
                    {(-i \omega + i q v_F)(-i\omega -iqv_F)}
\right] \\
&=& \frac{1}{4}\left[  K + K^{-1}-2  + \cos( 2 \omega x/ v )
(K^{-1}-K) \right] \;
.\end{eqnarray}
Hence the density of states at the boundary is found to vanish as
\begin{equation}
N(\epsilon) \sim |\epsilon|^{(1-K)/K}
\end{equation}
in agreement with the 
boundary exponent obtained in \cite{fabrizio95}, and the 
impurity exponent given in \cite{fabrizio1997}.
In fact this result has been considerably debated 
\cite{oreg1996,fabrizio1997,oreg1997}; compare also \cite{furusaki1997},
as well as the detailed presentation by von Delft and Schoeller 
\cite{delft1998}.\footnote{These authors emphasize the importance of
a proper handling of Klein factors, i.e.\ fermionic anticommutation
relations. For more recent studies of this aspect, see, for example, 
Ref.\ \cite{mocanu2004}.}

\section{Summary}
Quasiclassical theory is known to be a useful tool in the theoretical
description of superconductors and superfluids, including
non-equilibrium states and interfaces.
For normal-conducting electrons the theory mainly serves 
for the microscopic foundation
of a Boltzmann-like transport theory.
However, when taking into account the Coulomb interaction in terms of
a fluctuating Hubbard-Stratonovich field, the theory also captures 
important interaction effects, which are beyond the reach
of the Boltzmann equation. In this article we 
illustrated this by considering the one-dimensional version of the
theory, and demonstrated that it is possible to describe the
non-Fermi liquid physics of Luttinger liquids.
The mathematics involved is very similar to the $P(E)$ theory of
tunneling, which we made use of in this article, and also to the functional
bosonization approach to Luttinger liquids \cite{fogedby1976} 
(which we did not exploit here).
We concentrated on the spinless case and on thermal equilibrium;
the spin is straightforwardly included in the theory, and clearly --
since we use the Keldysh formalism -- non-equilibrium situations can be
covered as well. We are confident that in the future 
the quasiclassical theory will become a useful, complementary approach
to transport in interacting one-dimensional systems.
        
\begin{acknowledgement}
This work was supported by the Deutsche Forschungsgemeinschaft (SFB 484).
\end{acknowledgement}


\begin{thebibliography}{10}
\bibitem{schmid1975}A. Schmid and G. Sch\"on, 
  J. Low Temp. Phys. {\bf 20}, 207 (1975).

\bibitem{larkin1976}A. I. Larkin and Yu. N. Ovchinnikov,
  Sov. Phys. JETP {\bf 41}, 960 (1976);
  {\em ibid.} {\bf 46}, 155 (1977).

\bibitem{eckern1981}U. Eckern, Ann. Phys. (New York) {\bf 133}, 390 (1981).
  
\bibitem{eckern1981a}U. Eckern and A. Schmid, 
  J. Low Temp. Phys. {\bf 45}, 137 (1981).
   
\bibitem{shelankov2000}A. Shelankov and M. Ozana, 
   Phys. Rev. B {\bf 61}, 7077 (2000).

\bibitem{luck2001}T. L\"uck, U. Eckern, and A. Shelankov,
  Phys. Rev. B {\bf 63}, 064510 (2001).

\bibitem{luck2004}T. L\"uck and U. Eckern, 
   J. Phys.: Condens. Matter {\bf 16}, 2071 (2004).

\bibitem{raimondi2006}R. Raimondi, C. Gorini, P. Schwab, and M. Dzierzawa,
   Phys. Rev. B {\bf 74}, 035340 (2006).

\bibitem{schwab2006}P. Schwab, M. Dzierzawa, C. Gorini, and R. Raimondi, 
   preprint (2006), cond-mat/0606209.

\bibitem{kamenev1999}A. Kamenev and A. Andreev, 
  Phys. Rev. B {\bf 60}, 2218 (1999).

\bibitem{schwab2003}P. Schwab and R. Raimondi, 
   Ann. Phys. (Leipzig) {\bf 12}, 471 (2003).

\bibitem{tomonaga1950}S. Tomonaga, 
  Prog. Theor. Phys. {\bf 5}, 544 (1950).

\bibitem{luttinger1963}
J. M. Luttinger, J. Math. Phys. {\bf 4}, 1154 (1963).

\bibitem{haldane1981}F. D. M. Haldane,
  Phys. Rev. Lett. {\bf 47}, 1840 (1981).

\bibitem{keldysh1965}
L. V. Keldysh, Sov. Phys. JETP {\bf 20}, 1018 (1965).

\bibitem{rammer1986}
J. Rammer and H. Smith, Rev. Mod. Phys. {\bf 58}, 323 (1986).

\bibitem{ingold1992}
G.-L. Ingold and Yu. V. Nazarov, in: Single Charge Tunneling, edited by
H. Grabert and M. H. Devoret, NATO ASI Series, Series B: Physics, Vol.
294 (Plenum, New York, 1992), p. 21.

\bibitem{sassetti1994}
M. Sassetti and U. Weiss, Europhys. Lett. {\bf 27}, 311 (1994);
J. Rollb\"uhler and H. Grabert, Phys. Rev. Lett. {\bf 91}, 166402 (2003);
I. Safi and H. Saleur, Phys. Rev. Lett. {\bf 93}, 126602 (2004);
K. Le Hur and M.-R. Li, Phys. Rev. B {\bf 72}, 073305 (2005);
K. Le Hur, Phys. Rev. B {\bf 74}, 165104 (2006).

\bibitem{larkin1974}
I. E. Dzyaloshinskii and A. I. Larkin, Sov. Phys. JETP {\bf 38}, 202 (1974);
 \\ A. Luther and I. Peschel, Phys. Rev. B {\bf 9}, 2911 (1974).

\bibitem{oreg1996}Y.\ Oreg and A.\ M.\ Finkel'stein,
  Phys. Rev. Lett. {\bf 76}, 4230 (1996)

\bibitem{zaitsev1984}
A. V. Zaitsev, Sov. Phys. JETP {\bf 59}, 1015 (1984).

\bibitem{fabrizio95}
M. Fabrizio and A. Gogolin, Phys. Rev. B {\bf 51}, 17827 (1995);
S. Eggert, H. Johannesson, and A. Mattson, Phys. Rev. Lett. {\bf 76};
1505 (1996);
V. Meden, W. Metzner, U. Schollwock, O. Schneider, T. Stauber, and K.
Schonhammer, Eur. Phys. J. B {\bf 16}, 613 (2000).

\bibitem{fabrizio1997}M. Fabrizio and A. O. Gogolin,
  Phys. Rev. Lett. {\bf 78}, 4527 (1997).

\bibitem{oreg1997}
Y. Oreg and A. M. Finkel'stein, Phys. Rev. Lett. {\bf 78}, 4528 (1997).

\bibitem{furusaki1997}
A. Furusaki, Phys. Rev. B {\bf 56}, 9352 (1997).

\bibitem{delft1998}
J. von Delft and H. Schoeller, Ann. Phys. (Leipzig)
{\bf 7}, 225 (1998).

\bibitem{mocanu2004}
C. Mocanu, M. Dzierzawa, P. Schwab, and U. Eckern,
J. Phys.: Condens. Matter {\bf 16}, 6445 (2004);
phys. stat. sol. (b) {\bf 242}, 245 (2005).

\bibitem{fogedby1976} H. C. Fogedby,
 J. Phys. C: Solid State Phys. {\bf 9}, 3757 (1976);
 D. K. K. Lee and Y. Chen, J. Phys. A: Math. Gen. {\bf 21}, 4155 (1988);
 A. Grishin, I. V. Yurkevich, and I. V. Lerner, Phys. Rev. B {\bf 69}, 
 165108 (2004); I. V. Lerner and I. V. Yurkevich, Proceedings of LXXXI 
 Les Houches School on ``Nanoscopic quantum transport'', Les Houches, 
 France, June 28 -- July 30, 2004 [cond-mat/0508223].

\end{thebibliography}
\end{document}